\documentclass[12pt]{article}
\usepackage{graphicx}
\usepackage{amssymb}
\usepackage{mathptmx}
\usepackage[numbers]{natbib}
\usepackage{amssymb, amsmath, stmaryrd, latexsym, textcomp, wasysym,  ifsym}
\usepackage[unicode,pdfborder={0 0 0}]{hyperref}

\RequirePackage[a4paper]{geometry}
\begin{document}

\newcommand{\hdblarrow}{H\makebox[0.9ex][l]{$\downdownarrows$}-}
\date{}
\title{NMR Study of Disordered Inclusions in the Quenched Solid Helium}



\author{A.P. Birchenko, \and N.P. Mikhin, \and E.Ya.~Rudavskii, \and Ye.O. Vekhov}


\maketitle

\begin{center}
    \textit{B.Verkin Institute for Low Temperature Physics and Engineering of the National Academy of Sciences of Ukraine, 47 Lenin Ave., Kharkov 61103, Ukraine\\
yegor\_v@ukr.net}
\end{center}

\begin{abstract}
Phase structure of rapidly quenched solid helium samples is studied by the NMR technique. The pulse NMR method is used for measurements of spin-lattice $T_1$ and spin-spin $T_2$ relaxation times and spin diffusion coefficient $D$ for all coexisting phases. It was found that quenched samples are two-phase systems consisting of the hcp matrix and some inclusions which are characterized by $D$ and $T_2$ values close to those in liquid phase. Such liquid-like inclusions undergo a spontaneous transition to a new state with anomalously short $T_2$ times. It is found that inclusions observed in both the states disappear on careful annealing near the melting curve. It is assumed that the liquid-like inclusions transform into a new state --- a glass or a crystal with a large number of dislocations. These disordered inclusions may be responsible for the anomalous phenomena observed in supersolid region.
\end{abstract}

\section{Introduction}
The search for the supersolid state in solid helium has recently led to the detection of an anomalous behavior first in torsional experiments \cite{Kim.2004} and then in investigations of elastic properties \cite{Day.2007}, specific heat \cite{Lin.2007}, and mass transfer \cite{Ray.2010}. Although the effects observed are still waiting for a consistent explanation, the anomalies are most often credited to some type of disorder that can developed in a solid. For example, these maybe a system of dislocations, grain boundaries, liquid inclusions or a disordered (glassy) phase. To understand the phenomenon of supersolid, it is fundamentally important to clear up the conditions creating a disorder in  solid helium and the properties of the appearing disordered phase.

Much research has been carried out along this line employing various theoretical and experimental methods. The Monte-Carlo simulations \cite{Boninsegni.2006,Clark.2006,Pollet.2007} support the view that the supersolid state is impossible in a perfect hcp crystal of $^4$He. However, the signs of a disordered glassy phase were detected in torsion experiments investigating the relaxation dynamics \cite{Aoki.2008,Hunt.2009}. A large contribution of a disordered (glassy) phase to the pressure of $^4$He crystals grown on fast cooling \cite{Grigorev2.2007} and samples deformed $in\,situ$ \cite{Degtyarev.2010} was registered by precise barometry. The contribution of a glassy phase to the properties of solid helium was calculated in Refs.~\cite{Andreev.2007.ru,Balatsky.2007}. Interesting information was derived from visual observation of the disorder in solid $^4$He \cite{Sasaki.2008}: the samples grown by the blocking capillary method were polycrystals with grains of micrometer size and the liquid phase may well exist at the grain boundaries.

Much information about the phase composition of a sample can be obtained by the method of nuclear magnetic resonance (NMR). For example, the investigations of spin diffusion in a region of a bcc-hcp transition \cite{Mikhin.2001} revealed an additional diffusive process in a sample in which both crystallographic phases coexisted. The process is characterized by high diffusion coefficients and indicated that liquid-like inclusions can form in the sample in the course of the bcc-hcp transition. Later the presence of such inclusions in pure $^4$He and a dilute $^3$He-$^4$He solid mixtures was confirmed by precise pressure measurements \cite{Mikhin.2007}. The subsequent NMR investigations of the hcp phase \cite{Vekhov.2010} showed that regions with high diffusion coefficients appeared readily in fast-grown single-phase samples. It was found that in addition to the high diffusion coefficient, the inclusion had such spin-spin relaxation times which are inherent in the liquid phase \cite{Birchenko.2011.LT26}. The detected liquid-like inclusions disappeared after thorough annealing.

This study, which is a continuation of the previous NMR experiments \cite{Vekhov.2010,Birchenko.2011.LT26}, is concentrated on a detailed investigation of the revealed non-equilibrium phase. It is focused on identification of coexisting phases rather than the study of kinetics of the occurring processes.

Since NMR measurements in helium are usually made on $^3$He nuclei having a non-zero magnetic moment, a $^4$He crystal for this investigation must contain a certain amount of the $^3$He impurity so that technique used could be efficient. It is especially important for measurements in the region of supersolid effects (below $\AC 300$~mK) because the anomalies in question are sensitive even to low concentration of the $^3$He impurity. That is why the NMR experiments performed at very low temperatures \cite{Toda.2010,Kim.2010,Huan.2011} could furnish important information about the state of $^3$He in the supersolid region.

These NMR measurements were made at high temperatures (above $\AC 1.3$~K) to pursue another objective: to use $^3$He atoms as probes for identifying and investigating the metastable (disordered) phase in solid helium.

\section{Experimental Technique}

\begin{figure}[ht]
\begin{center}
\includegraphics[%
  width=0.5\linewidth,
  keepaspectratio]{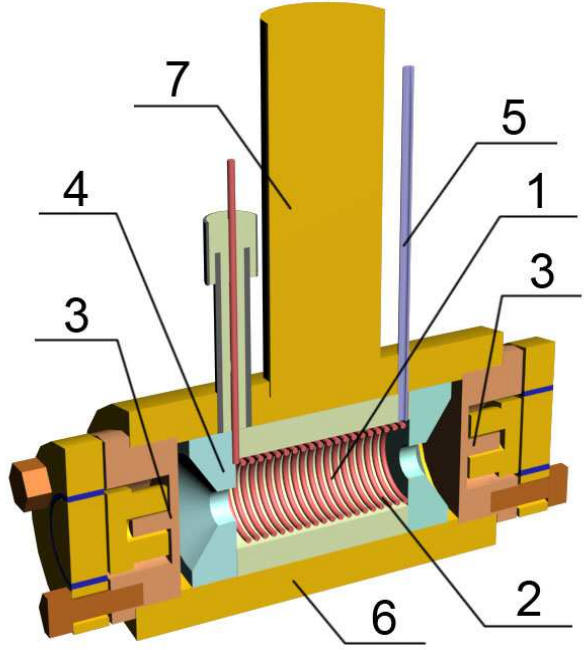}
\end{center}
\caption{(Color online) Schematic view of a NMR cell: 1 --- cell cavity; 2 --- NMR coil; 3 --- pressure gauges (BeCu); 4 --- fluoroplastic displacers; 5 --- cell filling capillary; 6 --- cell body (Cu); 7 --- cooling finger.}
\label{fig_cell}
\end{figure}

The experimental cell used for NMR measurements in solid helium is shown schematically in Fig.~\ref{fig_cell}.
The cell cavity (1) in a form of a cylinder 16 mm long and 8 mm in diameter was filled with helium. NMR coil (2) was wound in the inner surface of the cavity (sample). Straty-Adams pressure gauges (3) of $\pm 5-10$~mbar resolution was fixed at the cylinder ends. They were separated from the cylinder cavity with specially designed fluoroplastic displacers (4) to reduce edge effects on the NMR signal. The filling capillary (5) of the cell was made of stainless steel, its inner diameter being 0.1~mm. It was thermally connected to a 1K-pot to apply the blocking capillary method of growing solid helium samples. A resistance thermometer was mounted on the copper body (6) of the cell to measure the sample temperature (it had an accuracy of $\pm 5$~mK and a sensitivity of $\pm 1$~mK). The thermostabilizing system (heater and thermometer) attached in the upper part of a copper cooling finger (7) ensured temperature stabilization within $\pm 1$~mK. The systems of thermometry, thermostabilization, barometry, and NMR measurements were completely automated.

The measurements were made in the temperature range $1.3-2.0$~K. The hcp crystals under investigation corresponded to the pressures $34-40$~bar ($V_m=20.3-20.0$~cm$^3$/mol), i.e. were grown from the normal liquid above the upper triple point (bcc-hcp-He~I). The crystals were obtained by fast cooling along the melting curve at a rate of $\AC 2-6$~mK/s, which produced a large number of defects in the sample. Under such conditions the sample of solid helium contains also a large number of liquid-like inclusions that might be captured under crystallization (see Ref.~\cite{Birchenko.2011.LT26}).
 
The times of nuclear magnetic relaxation and the spin diffusion coefficients were measured in the NMR experiments. The measurements were made at the frequency $f_0=9.15$~MHz using sequences of probe pulses (the Carr-Purcell ($CP$) method) $90^o-\tau-180^o$ \cite{Carr.1954}, where $\tau$ is the time interval between the pulses.

The time of spin-spin relaxation $T_2$ and the spin diffusion coefficient $D$ were estimated by measuring the dependence of the relative echo-signal amplitude $h/h_0$ on the time interval $\tau$ between the probe pulses. In the $CP$ method, this dependence is \cite{Carr.1954}:
\begin{equation}
\frac{h}{h_0}=\sum_{i}\alpha_i\left[1-\exp \left( \frac{\Delta t}{(T_1)_i}\right)\right]\exp\left[-\frac{2\tau}{\left(T_2\right)_i}-\frac{2}{3}\gamma^2\tau^3G^2D_i\right] ,
\label{DCarr}
\end{equation} 
where $\Delta t$ is the time interval between the pulse sequences, $G$ is the magnetic field gradient, $(T_1)_i$ is the time of spin-lattice relaxation in the i-th phase, $(T_2)_i$ is the time of spin-spin relaxation in the i-th phase, $D_i$ and $\alpha_i$ are the diffusion coefficient and the relative volume content of the i-th phase, respectively; $\gamma$ is the gyromagnetic ratio. $T_1$, $T_2$, and $D$ can be found by varying the correspondent parameters in Eq.~(\ref{DCarr}). The time of spin-lattice relaxation $T_1$ can be obtained from the dependence $h/h_0(\Delta t)$ on varying $\Delta t$, the other parameters being constant. The spin-spin relaxation time $T_2$ can be obtained from the dependence $h/h_0(\tau)$ which is measured under the lowest magnetic field gradient ($G$ is close to zero) on varying $\tau$ and holding $\Delta t$ constant. Finally, the diffusion coefficient $D$ is derived from the dependence $h/h_0(\tau)$ under a non-zero magnetic field gradient with a invariable $\Delta t$.

If the system contains several subsystems (coexisting phases), Eq.~(\ref{DCarr}) gives a sum of the corresponding exponents and sought-for kinetic coefficients are obtainable from the analysis of the dependence obtained. The data obtained are reliable, which was proven by check NMR experiments in the single-phase regions of the mixture. They agree with available data for mixtures  $1.94\%$ $^3$He for relaxation times \cite{Miyoshi.1970,Guyer.1971}, $0.75\%$ and $2.17\%$ $^3$He for diffusion coefficients \cite{Grigorev.1973,Grigorev.1974} taking into account the difference in the $^3$He concentrations.

\section{Anomalous behavior of spin-spin relaxation time}
\label{T2}

The analysis of $T_1$-, $T_2$-, and $D$-values shows that the parameter $T_2$ (spin-spin relaxation time) and $D$ (diffusion coefficient) \cite{Vekhov.2010,Birchenko.2011.LT26} can provide much information necessary for identifying the phase structure of the sample. $T_2$ can tell much about the changes of the intensity of the spin motion. Such information was used to investigate the properties of solid helium in \cite{Allen.1982,Mikhin.2000}. According to the Bloembergen-Purcell-Pound (BPP) model \cite{Bloembergen.1948}, the rate of spin-spin relaxation decreases as the intensity of the relative motion of nuclei with a non-zero magnetic moment ($^3$He nuclei) increases. This correlation between $T_2$ and $D$ was quite fully considered in \cite{Cowan.1997.book}. On the other hand, when there is no additional strong magnetic effects (e.g. ferro- and diamagnetic impurities \cite{Korb.2010}) $T_2$, unlike $D$, is independent of the shape and the size of fine-grained objects. It is therefore appropriate to discuss the results obtained beginning with the time $T_2$.

\begin{figure}[h]
\begin{minipage}[t]{0.5\linewidth}
\includegraphics[width=1\linewidth]{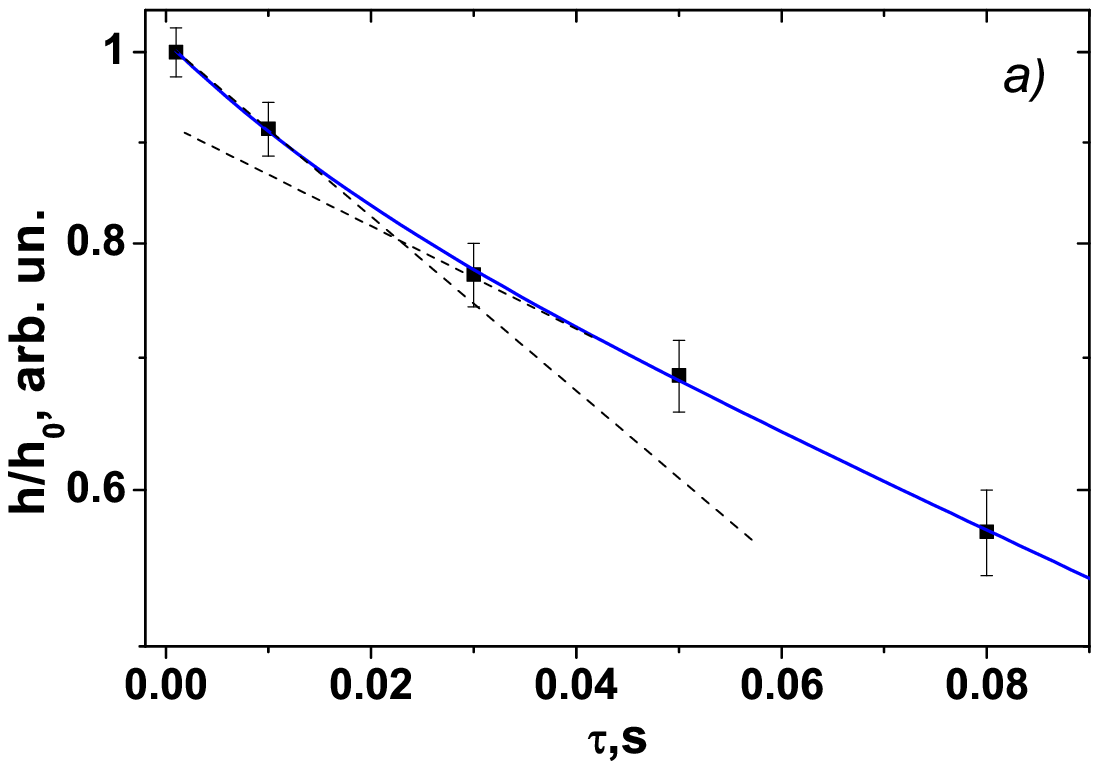}
\end{minipage}
\hfill 
\begin{minipage}[t]{0.5\linewidth}
\includegraphics[width=1\linewidth]{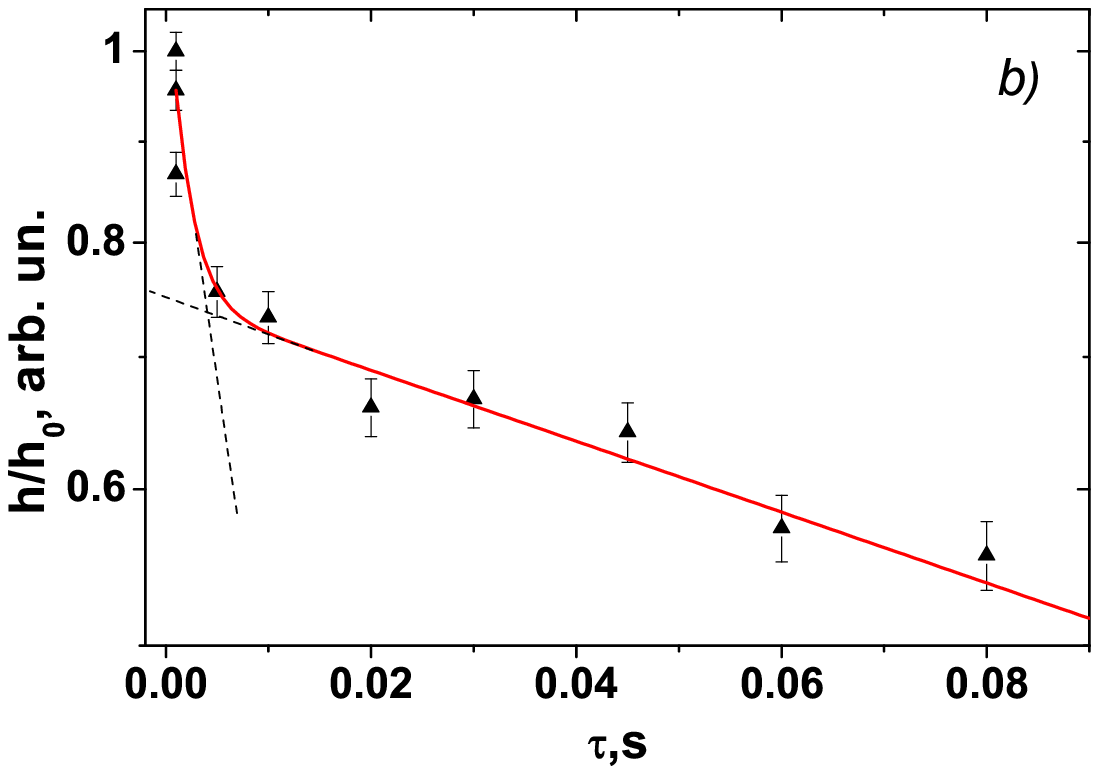}
\end{minipage}
\vfill 
\begin{minipage}[t]{0.5\linewidth}
\includegraphics[width=1\linewidth]{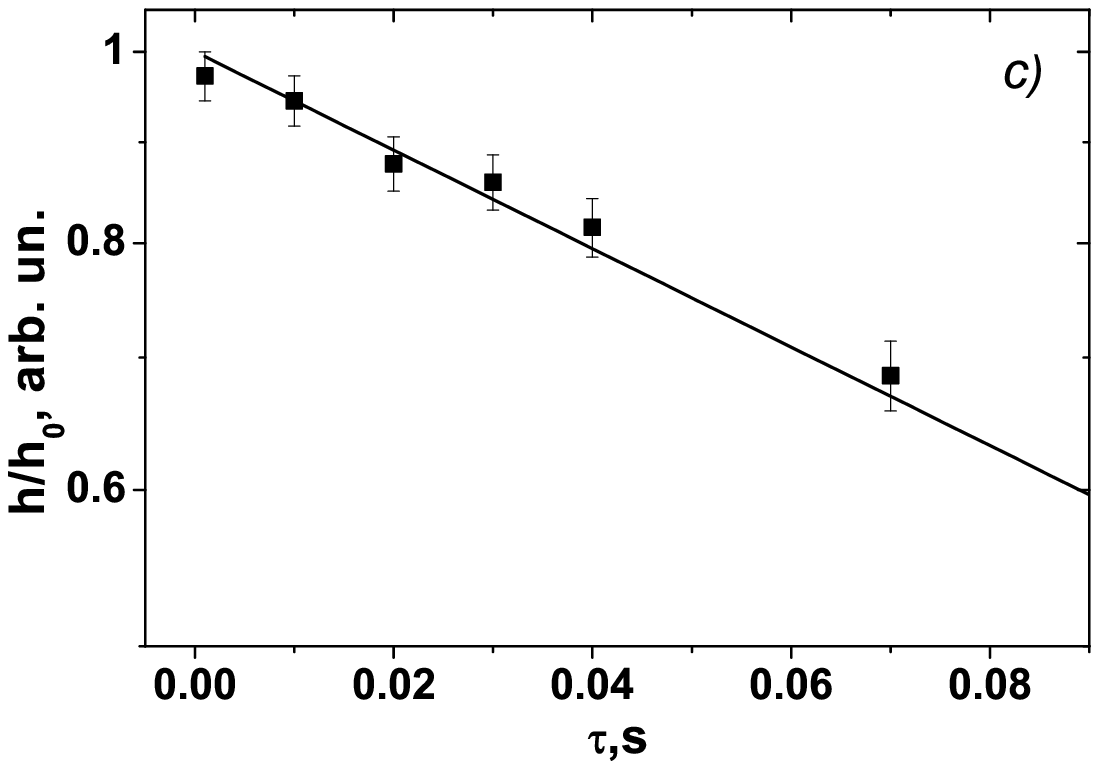}
\end{minipage}
\hspace{1pc}
\begin{minipage}[b]{18pc}
\caption{(Color online) The dependence of the spin-echo amplitude on the time interval between probe pulses ($G=0$) for different states of sample: a) as-grown fast-cooled sample, $T=1.7$~K, $P=35$~bar ($\Delta t=150$~s); b) the same sample after isothermal 3 hour exposure ($\Delta t=100$~s); c) the same sample after annealing near the melting curve and subsequent cooling to $T=1.7$~K ($\Delta t=100$~s). Dashed lines --- extrapolation of the dependence h/h0 for each phase.}
\label{fig_T2_hcp}
\end{minipage}
\end{figure}

The starting experimental dependences $h/h_0(\tau)$ taken on the same crystal in several states are illustrated in Fig.~\ref{fig_T2_hcp}. The as-grown hcp crystal obtained on cooling along the melting curve at the rate of $\AC 5$~mK/s is shown in Fig.~\ref{fig_T2_hcp}a. The dependence $h(\tau)$ has two slopes corresponding to two processes --- fast relaxation with the time $T_2=36\pm 21$~ms and slow relaxation with the time $T_2=362\pm 62$~ms. $T_2$ for the fast process measured at $T=1.7$~K is in good agreement with the spin-spin relaxation time for a bulk liquid at $P\AC 25$~bar; in the case of the slow process this time corresponds to the hcp phase. These data for the single-phase states of the investigating system were obtained in special calibration experiments \cite{Birchenko.2011.LT26}. They support our previous conclusion \cite{Vekhov.2010,Birchenko.2011.LT26} that metastable liquid-like inclusions are formed readily in fast grown crystals of solid helium.

An unexpected effect was observed in the course of NMR measurements when a fast grown sample with metastable liquid-like inclusions was investigated for about three hours at $T=1.7$~K. In this case the obtained dependence $h/h_0(\tau)$ is also a superposition of two exponents (Fig.~\ref{fig_T2_hcp}b), where the time $T_2$ for the hcp phase is now $443\pm 51$~ms (within the total experimental error it can be taken as practically unchanged). The time $T_2$ for the fast process connected with the formation of liquid-like inclusions decreased by an order of magnitude and was $3.2\pm 3.0$~ms. In all the cases this change in $T_2$ was rather fast: it occurred within one measurement run, i.e. no longer than $\Delta t\AC 100-300$~s. It is natural to assume that the liquid-like inclusions undergo some evolution or a phase transition in the hcp matrix.

The dependences processed by Eq.~(\ref{DCarr})  (Figs.~\ref{fig_T2_hcp}a and \ref{fig_T2_hcp}b) show that the weighting factor $\alpha_i$ for non-equilibrium inclusions (see Fig.~\ref{fig_T2_hcp}b) is about three times higher than in the case of Fig.~\ref{fig_T2_hcp}a (estimation of $\alpha_i$ is detailed in Section~\ref{discussion}).

\begin{figure}[ht]
\begin{center}
\includegraphics[%
  width=0.7\linewidth,
  keepaspectratio]{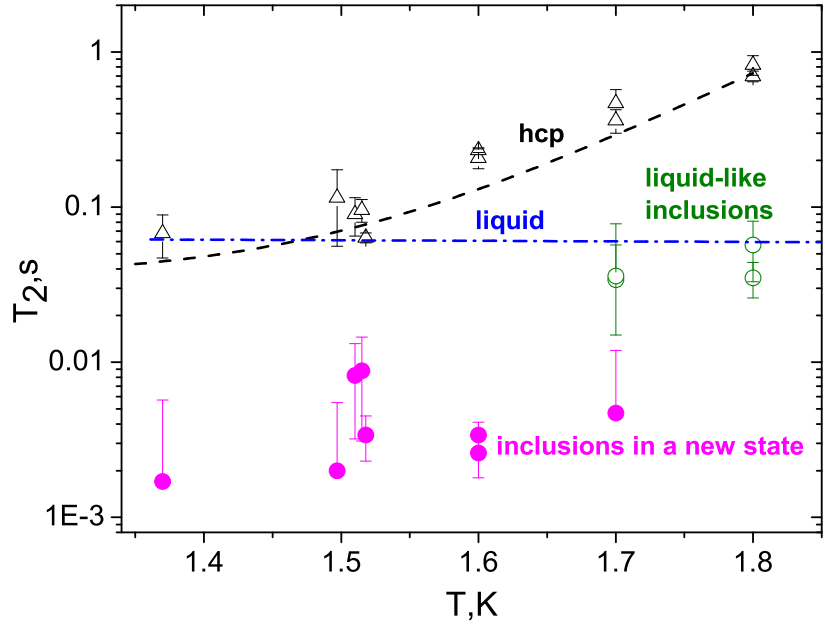}
\end{center}
\caption{(Color online) The temperature dependence of the spin-spin relaxation time $T_2$ for: (green $\Circle$) --- metastable liquid-like inclusions \cite{Birchenko.2011.LT26}, (red $\CIRCLE$) --- metastable inclusions after their transition to a new state, ($\triangle$) --- time $T_2$ for the hcp phase in the presence of metastable inclusions. Dashed line --- data for a single-phase hcp sample at $P\AC 35$~bar \cite{Birchenko.2011.LT26}, dash-dot line --- data for a bulk liquid at $P\AC 25$~bar \cite{Birchenko.2011.LT26}.
}
\label{fig_T2allinclusions}
\end{figure}

Note that the dependence $h(\tau)$ taken after thorough ($\AC 2$~hours) annealing of the sample near the melting curve ($T=1.95$~K) and its subsequent cooling to $T=1.7$~K can be described by one exponent with $T_2=350\pm 19$~ms, which is characteristic of the hcp phase at this temperature (see Fig.~\ref{fig_T2_hcp}c). This means that after annealing the non-equilibrium inclusions disappeared almost completely and the sample became a single-phase hcp crystal.

Similar results were also obtained at other temperatures (see Fig.~\ref{fig_T2allinclusions}). Fig.~\ref{fig_T2allinclusions} carries in addition the experimental data for the single-phase states of the system  the liquid state and the hcp phase (see Ref.~\cite{Birchenko.2011.LT26}). In the liquid state $T_2$ is practically temperature-independent (within the scatter of experimental data) in the region far from the degeneracy temperature. In the hcp phase $T_2$ decreases with lowering temperature and at $T\AC 1.5$~K the $T_2$-values of the crystal and the liquid almost coincide. The reason for this accidental coincidence is as follows: in the investigated temperature region the rate of spin-spin relaxation in the crystal is mainly determined by the concentration and the mobility of vacancies which intensify the motion of $^3$He impurity atoms. According to the BPP theory \cite{Bloembergen.1948}, the influence of the magnetic fields of the neighboring nuclear spins is averaged more efficiently and the spin-spin relaxation slows down. The Arrhenius equation suggests that $T_2$ decreases as the temperature of the crystal lowers. As a result, the contribution of the liquid-like inclusions is practically inseparable at $T<1.7$~K.

The $T_2$-values measured after the transition of the liquid-like inclusions to a new state are shown in Fig.~\ref{fig_T2allinclusions}. They are over the order of magnitude lower than $T_2$ of liquid-like inclusions. Note that the transition of liquid-like inclusions to a new state was observed in most of the investigated crystals with the delay time varying from half an hour to several hours. The process was spontaneous with no distinct correlation with temperature.

\section{Discussion}
\label{discussion}

The spin-spin relaxation time $T_2$ has been analyzed quite thoroughly (see the previous section). Unfortunately, the results obtained do not allow an unambiguous analysis of the spin-lattice relaxation time $T_1$. This is because $T_1$ is measured under operation of two mechanisms of relaxation in the liquid phase: the true relaxation is connected with recovery of equilibrium magnetization over the sample volume; a wall relaxation is determined by non-controllable interaction between $^3$He and the magnetic impurities at the cell walls. Although the experiment revealed some change in $T_1$ caused by the transition of the sample from the state with liquid-like inclusions to the new phase, an accurate quantitative analysis of the phase composition was problematic. Further experiments are therefore planned to investigate the behavior of the spin-lattice relaxation time in this system.

The behavior of the diffusion coefficient $D$ in samples grown on fast cooling is analyzed using the $\tau$-dependences of the spin echo-signal amplitude. These dependences are a superposition of two exponents \cite{Vekhov.2010,Birchenko.2011.LT26} --- a slowly attenuating exponent and a fast attenuating one which correspond to the hcp phase and the liquid-like inclusions, respectively. Note that the $D$-values measured for liquid-like inclusions are proportional to the rate of crystal growth \cite{Birchenko.2011.LT26}. This suggest that the size of the metastable liquid drops captured in the process of crystallization is the smaller for  the slower the crystallization rate. This may be due to a limited diffusion, like that usually observed when the size of liquid-like inclusions is comparable to the diffusion length $\sqrt{D\tau}$ (e.g., see Ref.~\cite{Wayne.1966}). In this case the coefficient $D$ is dependent on the size of the inclusions, $\tau$ being invariable.

The situation changed as the liquid-like inclusions transformed into a new state, with anomalously short times $T_2$. The echo-signal amplitudes has a wide scatter in data, which prohibited an accurate separation of the contribution made by the fast-diffusion process to the echo attenuation against the background of the hcp phase diffusion. However, the obtained dependence $h/h_0(\tau)$ could be described in terms of the slow-diffusion process typical of the hcp phase. This means that all of the liquid-like inclusions changed into the new state and the fast diffusion process did not observed. Yet, in some experiments the signs of the fast diffusion process were seen against the background of a wide data scatter, which suggest that some liquid-like inclusions did not transform into the new state. We may thus conclude that the diffusion coefficient of the new state is either smaller than $D$ of the hcp phase or it is of the same order of magnitude.

According to Eq.~(\ref{DCarr}), correct estimation of the relative contribution of $\alpha_i$ to the amplitudes of the echo-signal of the coexisting phases is possible if the time intervals $\Delta t$ between the sequences of probe pulses in NMR measurement are several times longer than the maximum time of spin-lattice relaxation in all phases, i.e. when magnetization is recovered completely. Only under this condition does the echo signal reach its maximum value. Assuming equal $^3$He concentrations in the phases, we can obtain the weighing factor $\alpha_i$ of the i-th phase.\footnote{The difference in the $^3$He concentration between crystalline and liquid phases is essential at temperatures below those used in this experiment.} Otherwise, magnetization of some phases can be recovered incomplete during NMR measurement because of the difference in the rates of spin-lattice relaxation in the phases, and this leads to a distortion of $\alpha_i$-values. This requirement is not always satisfied in real experiments because metastable inclusions evolve even during one run of $D$ or $T_2$ measurement. The obtained $\alpha_i$-values are therefore rather approximate. The estimation performed for annealed hcp samples using $\Delta t$ and $T_1$ shows that $\alpha_i$ of metastable inclusions can be as high as $\AC 20-30\%$.

The detected new phase is a long-living metastable state (see above) which disappears only after thorough annealing of the sample. Its properties differ from those of the hcp phase and liquid. We believe that the most probable reason for the decrease in $T_2$ of these inclusions is a transition of the liquid-like inclusions having higher diffusion coefficients to a new state with low $D$. The decrease in $D$ of the liquid entails a reduction of the relaxation time $T_2$ (see Section~\ref{T2}). This situation is similar to solidification of a liquid without the crystal order, i.e. the transition to an amorphous or glass-like state. Previously \cite{Menges.1991} the feasibility of formation of the amorphous phase in solid rare gases was found from heat capacity measurements on samples grown by condensation on a cold substrate. Unfortunately we are unaware of data on the relaxation time $T_2$ and the diffusion coefficient $D$ in amorphous helium, which makes the identification of the structure of the obtained phase somewhat uncertain. This assumption was however supported by precise measurements of pressure in solid $^4$He samples grown on fast cooling \cite{Grigorev2.2007}. Apart from the phonon influence, an additional contribution to the pressure $P_g$ proportional to $T^2$ was observed, which occurs typically in a disordered (glassy) phase. Recently, an amorphous helium phase has also been registered in neutron diffraction experiments \cite{Bossy.2010}. As follows from the structural factor analysis, liquid helium confined in a porous medium holds its liquid state to the pressure $P\AC 38$~bar at $T\AC 0.4$~K and then an amorphous phase forms when liquid helium solidifies. 
    
Crystallization of liquid-like inclusions generating a large number of dislocations can be another factor responsible for the lower $T_2$-values. It is known that  dislocations in crystals act as centers attracting impurities. In our case this attraction enriches the regions around dislocations in $^3$He atoms. The increasing local $^3$He concentration reduces the time of spin-spin relaxation in this region as $T_2 = 7.5\cdot 10^{-5}/x$~(s), where $x$ is the relative $^3$He concentration \cite{Mikhin.2000}. This dependence describes satisfactory other authors results \cite{Miyoshi.1970,Mikhin.2000,Greenberg.1972} and closely agrees with the calculation according to the Torrey theory \cite{Torrey.1953}.

There is one more scenario of a significant decrease in the time of nuclear magnetic relaxation \cite{Korb.2010}. Within a limited geometry the relaxation processes accelerate significantly in the presence of magnetic impurities on the walls. In this experiment a limited geometry could be created by crystalline dendrites growing into the volume of the liquid-like inclusions \cite{Franck.1986}, and the $^3$He atoms possessing nuclear paramagnetism could act as magnetic impurities. However, according to the Curie Law, the influence of such impurities is essential at rather low temperatures \cite{Mikhin.2004,Mikhin.2005.QFS2005}. It was therefore hardly probable in this experiment.

We also considered the possibility of the formation of bcc inclusions (in the $P-T$ phase diagram the bcc phase is quite close to the solid helium region under investigation). However, $T_2$ in the bcc phase is about  $\AC 0.1-1$~s \cite{Polturak.1995}, which is one or two orders of magnitude higher than the times of new state. 

Note that in the course of investigation on as-grown rapidly quenched samples both the pressure gauges in the cell exhibited a long-duration (up to 20~h) monotonic growth of pressure at a constant temperature. The rate of the pressure growth increased with the temperature, which suggests the thermally activated origin of the effect. The effect is attributed to the pressure gradient in the rapidly quenched samples. The pressure did not increase in samples grown carefully and slowly during $\AC 0.5-1$~h. The narrow areas in the displacers (see Fig.~\ref{fig_cell}) might considerably slow down the process of pressure equalizing. Therefore, we were unable to estimate pressure change due to annealing the metastable inclusions even having two pressure gauges.


\section{Conclusion}

The series of experiments in this study and Ref.~\cite{Birchenko.2011.LT26} have shown that a disordered metastable long living phase is readily formed in fast-grown helium crystals. The new phase coexists with the equilibrium crystalline phase. The measured diffusion coefficient and the spin-spin relaxation time of this phase correspond at first to the values typical for the liquid phase. The size of the liquid-like inclusions is the larger if the growth rate of the crystal is higher. The liquid-like inclusions do not form in crystals grown on cooling along the melting curve at comparatively low rates. The liquid-like inclusion disappear after thorough annealing near the melting temperature. They form at the stage of the crystal growth, and their size and quantity can be controlled by varying the growth rate of the crystal.

A new effect --- a spontaneous transition of liquid-like inclusions to another state  has been detected. The state has an anomalously short time of spin-spin relaxation unusual in both a crystal and a liquid. It is assumed that the new state is either an amorphous (glassy) phase or a crystal with a large number of dislocations. To identify the disordered phase definitely, further experimental investigations are necessary, in particular by structural methods.

In the context of the supersolid problem it is important to allow for the influence of the disordered phase on the properties of helium crystals. It is likely that the metastable disordered inclusions with very short times of spin-spin relaxation may be responsible for the anomalous phenomena in the region of supersolid.

The authors are indebted to V.A.~Maidanov, V.D.~Natsik, A.I.~Prokhvatilov, and D.A.~Tayurskiy for useful discussions while preparing this paper. The study was supported by the Science and Technology Center in Ukraine, Project \#5211.


\bibliographystyle{spphys}       
\bibliography{D:/vekhov/Thesis/My_Diss/bib_disser_vekhov}   

\begin{thebibliography}{10}
\providecommand{\url}[1]{{#1}}
\providecommand{\urlprefix}{URL }
\expandafter\ifx\csname urlstyle\endcsname\relax
  \providecommand{\doi}[1]{DOI \discretionary{}{}{}#1}\else
  \providecommand{\doi}{DOI \discretionary{}{}{}\begingroup
  \urlstyle{rm}\Url}\fi

\bibitem{Kim.2004}
E.~Kim, M.~Chan, Nature \textbf{427}, 225 (2004)

\bibitem{Day.2007}
J.~Day, J.~Beamish, Nature \textbf{450}, 853 (2007)

\bibitem{Lin.2007}
X.~Lin, A.~Clark, M.~Chan, Nature \textbf{449}, 1025 (2007)

\bibitem{Ray.2010}
M.~Ray, R.~Hallock, Phys. Rev. Lett. \textbf{105}(14), 145301 (2010)

\bibitem{Boninsegni.2006}
M.~Boninsegni, N.~Prokof'ev, B.~Svistunov, Phys. Rev. Lett. \textbf{96}(10),
  105301 (2006)

\bibitem{Clark.2006}
B.~Clark, D.~Ceperley, Phys. Rev. Lett. \textbf{96}(10), 105302 (2006)

\bibitem{Pollet.2007}
L.~Pollet, M.~Boninsegni, A.~Kuklov, N.~Prokof'ev, B.~Svistunov, M.~Troyer,
  Phys. Rev. Lett. \textbf{98}(13), 135301 (2007)

\bibitem{Aoki.2008}
Y.~Aoki, M.~Keiderling, H.~Kojima, Phys. Rev. Lett. \textbf{100}(21), 215303
  (2008)

\bibitem{Hunt.2009}
B.~Hunt, E.~Pratt, V.~Gadagkar, M.~Yamashita, A.~Balatsky, J.~Davis, Science
  \textbf{324}(5927), 632 (2009)

\bibitem{Grigorev2.2007}
V.~Grigor'ev, V.~Maidanov, V.~Rubanskii, S.~Rubets, E.~Rudavskii, A.~Rybalko,
  Y.~Syrnikov, V.~Tikhii, Phys. Rev. B \textbf{76}(22), 224524 (2007)

\bibitem{Degtyarev.2010}
I.~Degtyarev, A.~Lisunov, V.~Maidanov, V.~Rubanskiy, S.~Rubets, E.~Rudavskii,
  A.~Rybalko, V.~Tikhii, JETP \textbf{111}(4), 618 (2010)

\bibitem{Andreev.2007.ru}
A.~Andreev, JETP Letters \textbf{85}(11), 714 (2007)

\bibitem{Balatsky.2007}
A.~Balatsky, M.~Graf, Z.~Nussinov, S.~Trugman, Phys. Rev. B \textbf{75}(9),
  094201 (2007)

\bibitem{Sasaki.2008}
S.~Sasaki, F.~Caupin, S.~Balibar, JLTP \textbf{153}, 43 (2008)

\bibitem{Mikhin.2001}
N.~Mikhin, A.~Polev, E.~Rudavskii, JETP Letters \textbf{73}(9), 470 (2001)

\bibitem{Mikhin.2007}
N.~Mikhin, A.~Polev, E.~Rudavskii, Y.~Vekhov, JLTP \textbf{148}(5/6), 707
  (2007)

\bibitem{Vekhov.2010}
Y.~Vekhov, A.~Birchenko, N.~Mikhin, E.~Rudavskii, JLTP \textbf{158}, 496 (2010)

\bibitem{Birchenko.2011.LT26}
N.~Mikhin, A.~Birchenko, A.~Neoneta, E.~Rudavskii, Y.~Vekhov, J. Phys.: Conf.
  Series -- LT26  (to be published).
\newblock \urlprefix\url{http://arxiv.org/abs/1106.6135}

\bibitem{Toda.2010}
R.~Toda, P.~Gumann, K.~Kosaka, M.~Kanemoto, W.~Onoe, Y.~Sasaki,
  arXiv:1006.1025v1  (2010)

\bibitem{Kim.2010}
S.~Kim, C.~Huan, L.~Yin, J.~Xia, D.~Candela, N.~Sullivan, JLTP \textbf{158},
  584 (2010)

\bibitem{Huan.2011}
C.~Huan, S.~Kim, L.~Yin, J.~Xia, D.~Candela, N.~Sullivan, JLTP \textbf{162},
  167 (2011)

\bibitem{Carr.1954}
H.~Carr, E.~Purcell, Phys. Rev. \textbf{94}(3), 630 (1954)

\bibitem{Miyoshi.1970}
D.~Miyoshi, R.~Ñotts, R.~Greenberg, A.S. and.~Richardson, Phys. Rev. A
  \textbf{2}(3), 870 (1970)

\bibitem{Guyer.1971}
R.~Guyer, R.~Richardson, L.~Zane, Rev. Mod. Phys. \textbf{43}, 532 (1971)

\bibitem{Grigorev.1973}
V.~Grigor'ev, B.~Esel'son, V.~Mikheev, Sov. JETP \textbf{37}(2), 309 (1973)

\bibitem{Grigorev.1974}
V.~Grigor'ev, B.~Esel'son, V.~Mikheev, Sov. JETP \textbf{39}(1), 153 (1974)

\bibitem{Allen.1982}
A.~Allen, M.~Richards, J.~Schratter, JLTP \textbf{47}(3/4), 289 (1982)

\bibitem{Mikhin.2000}
N.~Mikhin, A.~Polev, E.~Rudavskii, Y.~Syrnikov, V.~Shvarts, Low Temp. Phys.
  \textbf{26}, 395 (2000)

\bibitem{Bloembergen.1948}
N.~Bloembergen, E.~Purcell, R.~Pound, Phys. Rev. \textbf{73}(7), 679 (1948)

\bibitem{Cowan.1997.book}
B.~Cowan, \emph{Nuclear magnetic resonance and relaxation} (Cambridge,
  University Press, 1997)

\bibitem{Korb.2010}
J.P. Korb, Comptes Rendus Physique \textbf{11}(2), 192  (2010)

\bibitem{Wayne.1966}
R.~Wayne, R.~Cotts, Phys. Rev. \textbf{151}(1), 264 (1966)

\bibitem{Menges.1991}
H.~Menges, H.~Lohneysen, JLTP \textbf{84}(3/4), 237 (1991)

\bibitem{Bossy.2010}
J.~Bossy, T.~Hansen, H.~Glyde, Phys. Rev. B \textbf{81}(18), 184507 (2010)

\bibitem{Greenberg.1972}
A.~Greenberg, W.~Thomlinson, R.~Richardson, JLTP \textbf{8}(1/2), 3 (1972)

\bibitem{Torrey.1953}
H.~Torrey, Phys. Rev. \textbf{92}(4), 962 (1953)

\bibitem{Franck.1986}
J.~Franck, J.~Jung, JLTP \textbf{64}(3/4), 165 (1986)

\bibitem{Mikhin.2004}
N.~Mikhin, Low Temp. Phys. \textbf{30}, 429 (2004)

\bibitem{Mikhin.2005.QFS2005}
N.~Mikhin, JLTP \textbf{138}(3/4), 817 (2005)

\bibitem{Polturak.1995}
E.~Polturak, I.~Schuster, I.~Berent, Y.~Carmi, S.~Lipson, B.~Chabaud, JLTP
  \textbf{101}(1/2), 177 (1995)

\end{thebibliography}

%
%

\end{document}